\documentclass[twocolumn]{aastex701}

\usepackage{graphicx}
\usepackage{amsmath}
\usepackage{hyperref}
\usepackage{longtable}
\usepackage{booktabs}

\newcommand{\te}{t_{\rm E}}
\newcommand{\thetae}{\theta_{\rm E}}

\newcommand{\dl}{D_{\rm L}}
\newcommand{\ds}{D_{\rm S}}
\def\e{{\rm E}}

\definecolor{brown}{rgb}{0.59, 0.29, 0.0}
\definecolor{darkgreen}{rgb}{0.0, 0.42, 0.24}
\definecolor{darkblue}{rgb}{0.01, 0.31, 0.59}
\definecolor{darkblue}{rgb}{0.0, 0.25, 0.42}
\definecolor{blue}{rgb}{0.0,0.0,1.0}
\definecolor{green}{rgb}{0.0,1.0,0.0}

\begin{document}

\title{Four-Body Gravitational Microlensing Events Involving Both a Binary Lens and a Binary Source}
\shorttitle{ Microlensing Events Involving a Binary Lens and a Binary Source}

\author{Cheongho Han}
\affiliation{Department of Physics, Chungbuk National University, Cheongju 28644, Republic of Korea}
\email{cheongho@astroph.chungbuk.ac.kr}
\author{Chung-Uk Lee}
\affiliation{Korea Astronomy and Space Science Institute, Daejon 34055, Republic of Korea}
\email{leecu@kasi.re.kr}
\author{Andrzej Udalski} 
\affiliation{Astronomical Observatory, University of Warsaw, Al.~Ujazdowskie 4, 00-478 Warszawa, Poland}
\email{udalski@astrouw.edu.pl} 
\collaboration{14}{(Leading authors)}
\author{Michael D. Albrow}   
\affiliation{University of Canterbury, Department of Physics and Astronomy, Private Bag 4800, Christchurch 8020, New Zealand}
\email{michael.albrow@canterbury.ac.nz}
\author{Sun-Ju Chung}
\affiliation{Korea Astronomy and Space Science Institute, Daejon 34055, Republic of Korea}
\email{sjchung@kasi.re.kr}
\author{Andrew Gould}
\affiliation{Department of Astronomy, Ohio State University, 140 West 18th Ave., Columbus, OH 43210, USA}
\email{gould.34@osu.edu}
\author{Youn Kil Jung}
\affiliation{Korea Astronomy and Space Science Institute, Daejon 34055, Republic of Korea}
\affiliation{University of Science and Technology, Daejeon 34113, Republic of Korea}
\email{younkil21@gmail.com}
\author{Kyu-Ha~Hwang}
\affiliation{Korea Astronomy and Space Science Institute, Daejon 34055, Republic of Korea}
\email{kyuha@kasi.re.kr}
\author{Yoon-Hyun Ryu}
\affiliation{Korea Astronomy and Space Science Institute, Daejon 34055, Republic of Korea}
\email{yhryu@kasi.re.kr}
\author{Yossi Shvartzvald}
\affiliation{Department of Particle Physics and Astrophysics, Weizmann Institute of Science, Rehovot 76100, Israel}
\email{yossishv@gmail.com}
\author{In-Gu Shin}
\affiliation{Department of Astronomy, Westlake University, Hangzhou 310030, Zhejiang Province, China}
\email{ingushin@gmail.com}
\author{Jennifer C. Yee}
\affiliation{Center for Astrophysics $|$ Harvard \& Smithsonian 60 Garden St., Cambridge, MA 02138, USA}
\email{jyee@cfa.harvard.edu}
\author{Weicheng Zang}
\affiliation{Department of Astronomy, Westlake University, Hangzhou 310030, Zhejiang Province, China}
\email{zangweicheng@westlake.edu.cn}
\author{Hongjing Yang}
\affiliation{Department of Astronomy, Westlake University, Hangzhou 310030, Zhejiang Province, China}
\email{yanghongjing@westlake.edu.cn}
\author{Doeon Kim}
\affiliation{Department of Physics, Chungbuk National University, Cheongju 28644, Republic of Korea}
\email{qso21@hanmail.net}
\author{Dong-Jin Kim}
\affiliation{Korea Astronomy and Space Science Institute, Daejon 34055, Republic of Korea}
\email{keaton03@kasi.re.kr}
\author{Seung-Lee Kim}
\affiliation{Korea Astronomy and Space Science Institute, Daejon 34055, Republic of Korea}
\email{slkim@kasi.re.kr}
\author{Dong-Joo Lee}
\affiliation{Korea Astronomy and Space Science Institute, Daejon 34055, Republic of Korea}
\email{marin678@kasi.re.kr}
\author{Sang-Mok Cha}
\affiliation{Korea Astronomy and Space Science Institute, Daejon 34055, Republic of Korea}
\email{chasm@kasi.re.kr}
\author{Yongseok Lee}
\affiliation{Korea Astronomy and Space Science Institute, Daejon 34055, Republic of Korea}
\email{yslee@kasi.re.kr}
\author{Byeong-Gon Park}
\affiliation{Korea Astronomy and Space Science Institute, Daejon 34055, Republic of Korea}
\email{bgpark@kasi.re.kr}
\author{Richard W. Pogge}
\affiliation{Department of Astronomy, Ohio State University, 140 West 18th Ave., Columbus, OH 43210, USA}
\email{pogge.1@osu.edu}
\collaboration{20}{(KMTNet Collaboration)}
\author{Przemek Mr{\'o}z}
\affiliation{Astronomical Observatory, University of Warsaw, Al.~Ujazdowskie 4, 00-478 Warszawa, Poland}
\email{pmroz@astrouw.edu.pl}
\author{Micha{\l} K. Szyma{\'n}ski}
\affiliation{Astronomical Observatory, University of Warsaw, Al.~Ujazdowskie 4, 00-478 Warszawa, Poland}
\email{msz@astrouw.edu.pl}
\author{Jan Skowron}
\affiliation{Astronomical Observatory, University of Warsaw, Al.~Ujazdowskie 4, 00-478 Warszawa, Poland}
\email{jskowron@astrouw.edu.pl}
\author{Rados{\l}aw Poleski} 
\affiliation{Astronomical Observatory, University of Warsaw, Al.~Ujazdowskie 4, 00-478 Warszawa, Poland}
\email{radek.poleski@gmail.co}
\author{Igor Soszy{\'n}ski}
\affiliation{Astronomical Observatory, University of Warsaw, Al.~Ujazdowskie 4, 00-478 Warszawa, Poland}
\email{soszynsk@astrouw.edu.pl}
\author{Pawe{\l} Pietrukowicz}
\affiliation{Astronomical Observatory, University of Warsaw, Al.~Ujazdowskie 4, 00-478 Warszawa, Poland}
\email{pietruk@astrouw.edu.pl}
\author{Szymon Koz{\l}owski} 
\affiliation{Astronomical Observatory, University of Warsaw, Al.~Ujazdowskie 4, 00-478 Warszawa, Poland}
\email{simkoz@astrouw.edu.pl}
\author{Krzysztof A. Rybicki}
\affiliation{Astronomical Observatory, University of Warsaw, Al.~Ujazdowskie 4, 00-478 Warszawa, Poland}
\email{krybicki@astrouw.edu.pl}
\author{Patryk Iwanek}
\affiliation{Astronomical Observatory, University of Warsaw, Al.~Ujazdowskie 4, 00-478 Warszawa, Poland}
\email{piwanek@astrouw.edu.pl}
\author{Krzysztof Ulaczyk}
\affiliation{Department of Physics, University of Warwick, Gibbet Hill Road, Coventry, CV4 7AL, UK}
\email{kulaczyk@astrouw.edu.pl}
\author{Marcin Wrona}
\affiliation{Astronomical Observatory, University of Warsaw, Al.~Ujazdowskie 4, 00-478 Warszawa, Poland}
\affiliation{Villanova University, Department of Astrophysics and Planetary Sciences, 800 Lancaster Ave., Villanova, PA 19085, USA}
\email{mwrona@astrouw.edu.pl}
\author{Mariusz Gromadzki}          
\affiliation{Astronomical Observatory, University of Warsaw, Al.~Ujazdowskie 4, 00-478 Warszawa, Poland}
\email{marg@astrouw.edu.pl}
\author{Mateusz J. Mr{\'o}z} 
\affiliation{Astronomical Observatory, University of Warsaw, Al.~Ujazdowskie 4, 00-478 Warszawa, Poland}
\email{mmroz@astrouw.edu.pl}
\collaboration{100}{(The OGLE Team)}
\correspondingauthor{\texttt{leecu@kasi.re.kr}}

\begin{abstract}
We present detailed analyses of three anomalous microlensing events--KMT-2021-BLG-0209, 
KMT-2021-BLG-0901, and OGLE-2025-BLG-0356--identified from a systematic re-examination 
of KMTNet light curves for which previous modeling attempts failed or left persistent 
residuals. Although all three events show caustic-related features consistent with 
binary-lens microlensing, we find that their full light-curve structures 
can be described by four-body configurations that required
four-body configurations involving a binary lens and a binary source. In KMT-2021-BLG-0209, 
weak caustic-exit residuals arise from a faint companion source undergoing an additional 
caustic interaction. In KMT-2021-BLG-0901, a late-time re-brightening is produced when 
the secondary source encounters the resonant caustic long after the primary. For 
OGLE-2025-BLG-0356, we test the degeneracy between 3L1S and 2L2S interpretations of a 
short isolated anomaly and find that the 2L2S model provides a significantly better fit.
Source colors and magnitudes indicate binary sources composed of (G8V, M3V), (G8V, K2V), 
and (G6V, G8V) stars for the three events, respectively. Bayesian inference suggests that 
the lenses are predominantly low-mass binaries, including one system (KMT-2021-BLG-0901) 
with a companion consistent with a brown dwarf. These events add to the growing sample of 
well-characterized 2L2S systems and underscore the importance of systematically testing 
complex models, particularly in anticipation of the high-precision microlensing data 
expected from the Roman Space Telescope survey. 
\end{abstract}


\section{Introduction \label{sec:one}} 

Given that most stars reside in binary or higher-order multiple systems \citep{Duquennoy1991, 
Raghavan2010}, microlensing events involving both a binary lens and a binary source (2L2S) 
should, in principle, occur frequently.  Nevertheless, reports of confirmed 2L2S events 
were rare during the era of first-generation microlensing surveys such as OGLE-I, MACHO, 
and EROS \citep{Udalski1992, Alcock1993, Aubourg1993}. This scarcity was due in part to 
limited temporal sampling, but also to the intrinsic complexity of 2L2S light curves, which 
made their interpretation challenging and often led to ambiguous or incomplete modeling.

As a result, 2L2S lensing events were long regarded as unusual and largely anecdotal. 
Recent high-cadence microlensing surveys, including OGLE-IV, MOA, and KMTNet 
\citep{Udalski2015, Sumi2013, Kim2016}, have fundamentally changed this situation.  
Continuous, high-frequency monitoring of dense stellar fields has demonstrated that 
2L2S events arise naturally in microlensing observations and are not pathological 
exceptions. Their previously perceived rarity was therefore primarily the consequence 
of observational limitations and interpretive challenges, rather than an intrinsically 
low occurrence rate.

\begin{deluxetable}{lllllll}
\tablewidth{0pt}
\tablecaption{Coordinates and event ID correspondence. \label{table:one}}
\tablehead{
\multicolumn{1}{c}{KMTNet}                    &
\multicolumn{1}{c}{(RA, DEC)$_{\rm J2000}$}   &
\multicolumn{1}{c}{$(l, b)$}                  &
\multicolumn{1}{c}{Other ID}                       
}
\startdata
 KMT-2021-BLG-0209   &  (17:52:49.81, $-30$:22:40.19)  &  ($-0^\circ$\hskip-2pt.4307, $-2^\circ$\hskip-2pt.0848)   &  \nodata            \\
 KMT-2021-BLG-0901   &  (17:39:45.77, $-28$:11:35.92)  &  ($-0^\circ$\hskip-2pt.0451, $+1^\circ$\hskip-2pt.4845)   &  \nodata            \\
 OGLE-2025-BLG-0356  &  (17:51:57.59, $-30$:23:40.20)  &  ($-0^\circ$\hskip-2pt.5405, $-1^\circ$\hskip-2pt.9316)   &  KMT-2025-BLG-0729  \\
\enddata
\end{deluxetable}

As the number of securely identified 2L2S events has increased, the scientific emphasis 
has shifted from the presentation of isolated curiosities to a comparative understanding 
of these systems as a class. Individual 2L2S events are information-rich, but their 
broader significance emerges most clearly when multiple events are analyzed together. 
Comparative studies reveal recurring light-curve morphologies, common modeling challenges, 
and systematic degeneracies that are difficult to recognize from single-event analyses 
alone. Publishing additional well-characterized 2L2S events thus plays an essential 
role in establishing a reference sample for interpreting complex microlensing anomalies.

The importance of 2L2S events is further heightened in the era of forthcoming space-based 
microlensing surveys. Missions such as the Nancy Grace Roman Space Telescope 
\citep{Spergel2015, Penny2019} will routinely deliver high-precision and continuous
light curves, in which subtle deviations from standard microlensing models are expected 
to be common. In this observational regime, multiplicity in either the source or the 
lens can leave measurable imprints on the light curve and, if unrecognized, may lead to 
ambiguous or incorrect physical interpretations. Well-characterized 2L2S events from 
current ground-based surveys therefore provide critical test cases for validating modeling 
strategies, anomaly-classification schemes, and end-to-end analysis pipelines that will 
ultimately be applied to Roman data.

In this context, the motivation for publishing additional 2L2S events is not to infer 
event rates or population-level statistics, but rather to strengthen the empirical 
foundation needed to interpret the diversity and consequences of complex microlensing 
configurations. Each new event contributes to a growing body of evidence that multiplicity 
in the lens and/or source is a fundamental aspect of microlensing analysis--not merely an 
occasional complication--and that properly accounting for it is essential for robust 
stellar and exoplanet microlensing studies.

In this paper, we report three newly identified 2L2S microlensing events: KMT-2021-BLG-0209, 
KMT-2021-BLG-0901, and OGLE-2025-BLG-0356.  These events were discovered as part of a 
systematic effort to investigate anomalous microlensing light curves detected by the 
KMTNet survey for which no satisfactory interpretations had previously been reported. 
The first two 2L2S events uncovered in this program (OGLE-2018-BLG-0584 and KMT-2018-BLG-2119) 
were analyzed by \citet{Han2023}. Subsequently, \citet{Han2024} presented analyses of 
three additional events (KMT-2021-BLG-0284, KMT-2022-BLG-2480, and KMT-2024-BLG-0412), 
while \citet{Han2025b} reported another three events (OGLE-2024-BLG-0657, KMT-2024-BLG-2017, 
and KMT-2024-BLG-2480). Here we provide detailed analyses of three newly discovered events.

\section{Observation and data \label{sec:two}}

The coordinates and identification references for the three 2L2S events are listed 
in Table~\ref{table:one}. Of these, KMT-2021-BLG-0209 and KMT-2021-BLG-0901 were 
observed exclusively by the KMTNet survey. OGLE-2025-BLG-0356, on the other hand, 
was observed by both the OGLE and KMTNet surveys, and is also designated as 
KMT-2025-BLG-0729 in the KMTNet catalog. Because the lensing-induced magnification 
of this event was detected earlier by OGLE than by KMTNet, we adopt the OGLE 
identification throughout this paper.

The photometric data used in this analysis were obtained from the KMTNet survey for 
all three events, supplemented by OGLE data for OGLE-2025-BLG-0356.  KMTNet operates 
three identical 1.6 m telescopes located in Chile (KMTC), South Africa (KMTS), and 
Australia (KMTA), providing near-continuous coverage of the Galactic bulge 
\citep{Kim2016}.  Each telescope is equipped with a wide-field mosaic CCD camera 
with a field of view of 4 deg$^2$ and a pixel scale of 0.4 arcsec/pixel.  OGLE-IV 
uses a 1.3 m telescope located in Chile, equipped with a mosaic CCD camera covering 
1.4 deg$^2$ with a pixel scale of 0.26 arcsec per pixel \citep{Udalski2015}.  Both 
surveys observe primarily in the Cousins $I$ band, with occasional $V$-band observations 
for color measurements. The photometry was derived using pipelines developed by the 
individual survey teams, namely the pipeline described by \citet{Albrow2009} for the 
KMTNet survey and that of \citet{Udalski2003} for the OGLE survey. Both pipelines are 
based on difference image analysis optimized for crowded stellar fields \citep{Tomaney1996, 
Alard1998, Wozniak2000}.

Prior to modeling, the KMTNet data were re-reduced using the updated photometry 
pipeline of \citet{Yang2024}. The resulting light curves were cleaned by iteratively 
removing outliers caused by poor observing conditions, instrumental artifacts, or 
reduction failures, and then normalized to ensure consistency among data sets. 
Because difference image analysis often yields underestimated or inconsistent 
uncertainties, we renormalized the error bars for each data set to obtain 
statistically consistent fits, rescaling them such that the $\chi^2$ per degree of 
freedom is unity.  Specifically, we applied
\begin{equation}
\sigma' = k \sqrt{\sigma^2 + \sigma_{\rm min}^2},
\label{eq1}
\end{equation}
\hskip-4pt
where $\sigma$ is the original uncertainty, $k$ is a scaling factor, and $\sigma_{\rm min}$ 
is a minimum error floor that accounts for systematic effects. The parameters $k$ and 
$e_{\rm min}$ were determined separately for each data set by requiring the cumulative 
$\chi^2$ distribution as a function of magnification to be approximately linear 
\citep{Yee2012}.  In Table~\ref{table:two}, we list the error-bar rescaling factors 
applied to the individual data sets for each event.

\begin{deluxetable}{lllllll}
\tablewidth{0pt}
\tablecaption{Error bar rescling factors. \label{table:two}}
\tablehead{
\multicolumn{1}{c}{Data set}             &
\multicolumn{1}{c}{$k$}                  &
\multicolumn{1}{c}{$\sigma_{\rm min}$}                       
}
\startdata
 KMT-2021-BLG-0209   &        &         \\
 \hskip4pt KMTC01    & 1.187  & 0.01    \\    
 \hskip4pt KMTC41    & 1.029  & 0.03    \\    
 \hskip4pt KMTS01    & 0.970  & 0.03    \\    
 \hskip4pt KMTS41    & 1.015  & 0.03    \\    
 \hskip4pt KMTA01    & 0.985  & 0.01    \\    
 \hskip4pt KMTA41    & 1.140  & 0.01    \\    
\hline
 KMT-2021-BLG-0901   &        &         \\
 \hskip4pt KMTC14    & 1.072  & 0.02    \\    
 \hskip4pt KMTS14    & 1.183  & 0.02    \\    
 \hskip4pt KMTA14    & 1.086  & 0.02    \\    
\hline
 OGLE-2025-BLG-0356  &        &         \\
 \hskip4pt KMTC01    & 1.025  & 0.02    \\    
 \hskip4pt KMTC41    & 0.990  & 0.02    \\    
 \hskip4pt KMTS01    & 1.000  & 0.02    \\    
 \hskip4pt KMTS41    & 0.975  & 0.02    \\    
 \hskip4pt KMTA01    & 0.806  & 0.02    \\    
 \hskip4pt KMTA41    & 0.917  & 0.02    \\  
 \hskip4pt OGLE      & 1.476  & 0.02    \\    
\enddata
\end{deluxetable}

\begin{figure*}[t]
\centering
\includegraphics[width=13.0cm]{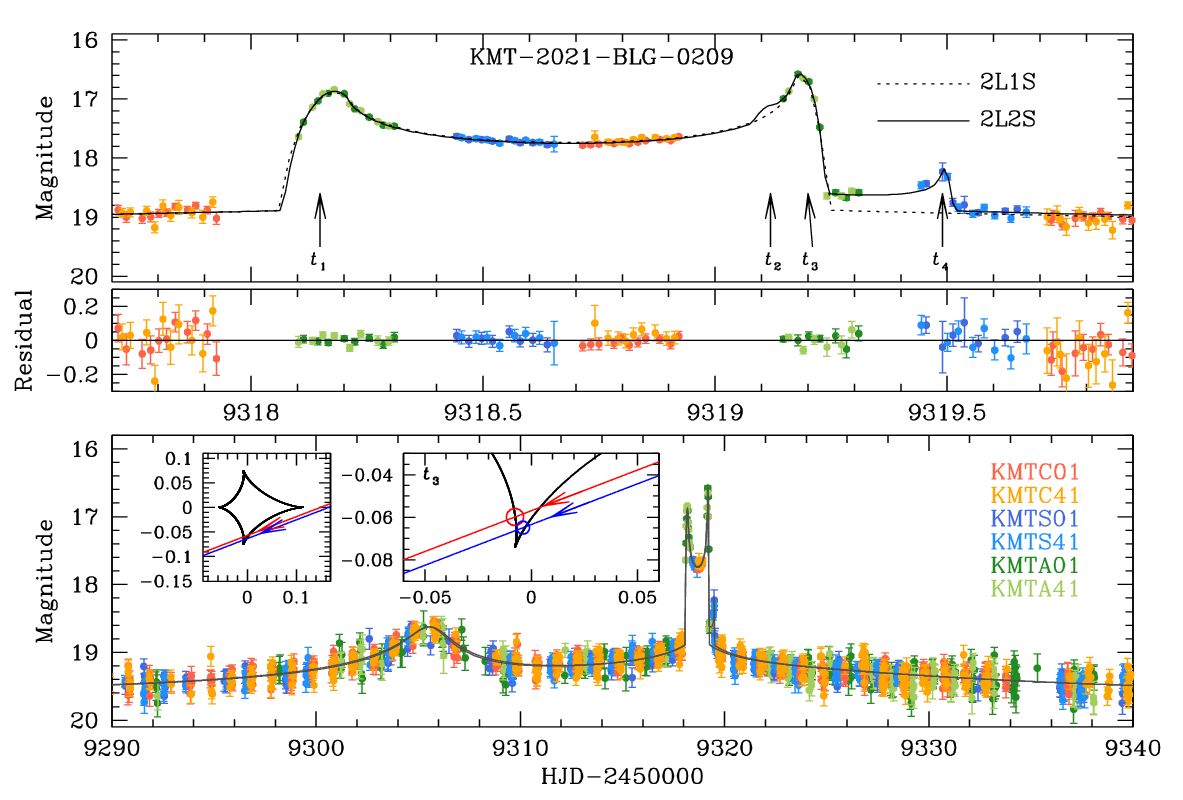}
\caption{
Light curve of the microlensing event KMT-2021-BLG-0209. The lower panel shows 
the full light curve, while the upper panel presents a zoomed-in view of the 
caustic-crossing region. The dotted and solid curves represent the best-fit models 
based on the 2L1S and 2L2S configurations, respectively. The two insets in the lower 
panel display the source trajectories relative to the caustic structure. The left 
inset shows the overall caustic geometry, and the right inset provides an enlarged 
view of the region near the caustic crossings. The arrowed trajectories of the 
primary and secondary source stars are indicated in red and blue, respectively.
The two small circles on the source trajectories indicate the positions of $S_1$ 
and $S_2$ at time $t_3$, as marked in the upper panel. Because the scaled source 
size is too small to be clearly visible, the circles are not drawn to scale.
}
\label{fig:one}
\end{figure*}

\section{Light curve analysis \label{sec:three}}

The light curves of all three events exhibit prominent caustic-crossing features, 
for which a multi-component lensing configuration is the only known cause.  We 
therefore begin our analysis by modeling the light curves with a binary-lens 
single-source (2L1S) configuration.

\begin{deluxetable}{lllllll}
\tablewidth{0pt}
\tablecaption{Best-fit lensing parameters. \label{table:three}}
\tablehead{
\multicolumn{1}{c}{Parameter}           &
\multicolumn{1}{c}{KMT-2021-BLG-0209 }  &
\multicolumn{1}{c}{KMT-2021-BLG-0901 }  &
\multicolumn{2}{c}{OGLE-2025-BLG-0356 } \\
\multicolumn{1}{c}{ }                   &
\multicolumn{1}{c}{ }                   &
\multicolumn{1}{c}{ }                   &
\multicolumn{1}{c}{2L2S}                &
\multicolumn{1}{c}{3L1S}   
}
\startdata
 $\chi^2$                  &  $3406.6              $  &  $1770.7              $   &  $4829.0           $  &  $4936.5              $   \\
 $t_{0,1}$ (HJD$^\prime$)  &  $9316.790 \pm 0.026  $  &  $9368.358 \pm 0.054  $   &  $766.45 \pm 0.31  $  &  $763.80 \pm 0.24     $   \\
 $u_{0,1}$                 &  $0.0531 \pm 0.0008   $  &  $0.0635 \pm 0.0012   $   &  $0.5580 \pm 0.0064$  &  $0.5171 \pm 0.0056   $   \\
 $t_{0,2}$ (HJD$^\prime$)  &  $9316.937 \pm 0.030  $  &  $9511.091 \pm 4.163  $   &  $757.41 \pm 0.31  $  &  \nodata                  \\
 $u_{0,2}$                 &  $0.0592 \pm 0.0010   $  &  $0.190 \pm 0.070     $   &  $0.6372 \pm 0.0062$  &  \nodata                  \\
 $\te$ (days)              &  $84.94 \pm 1.63      $  &  $41.38 \pm 0.94      $   &  $33.27 \pm 0.14   $  &  $31.01 \pm 0.33      $   \\
 $s$                       &  $3.18 \pm 0.024      $  &  $1.1489 \pm 0.0054   $   &  $0.7222 \pm 0.0065$  &  $0.6628 \pm 0.0044   $   \\
 $q$                       &  $1.346 \pm 0.066     $  &  $0.1058 \pm 0.0055   $   &  $1.997 \pm 0.093  $  &  $2.793 \pm 0.060     $   \\
 $\alpha$ (rad)            &  $5.9172 \pm 0.0029   $  &  $0.611 \pm 0.014     $   &  $5.035 \pm 0.017  $  &  $4.813 \pm 0.008     $   \\
 $\rho_1$ ($10^{-3}$)      &  $0.42 \pm 0.01       $  &  $1.67 \pm 0.22       $   &  $2.28 \pm 0.10    $  &  $0.0027 \pm 0.0001   $   \\
 $\rho_2$ ($10^{-3}$)      &  \nodata                 &  \nodata                  &  $1.09 \pm 0.13    $  &  \nodata                  \\
 $q_F$                     &  $0.0745 \pm 0.0059   $  &  $0.43 \pm 0.20       $   &  $0.529 \pm 0.015  $  &  \nodata                  \\
 $s_3$                     &  \nodata                 &  \nodata                  &  \nodata              &  $1.7806 \pm 0.0075   $   \\
 $q_3$ ($10^{-3}$)         &  \nodata                 &  \nodata                  &  \nodata              &  $7.96 \pm 0.99       $   \\
 $\psi$ (rad)              &  \nodata                 &  \nodata                  &  \nodata              &  $5.4654 \pm 0.007    $   \\
\enddata
\tablecomments{
Here ${\rm HJD}^\prime = {\rm HJD}-2450000$ for KMT-2021-BLG-0209 and KMT-2021-BLG-0901, and ${\rm HJD}^\prime = {\rm HJD}-2460000$ for OGLE-2025-BLG-0356.
}
\end{deluxetable}

In the 2L1S framework, the lensing light curve is described by seven fundamental parameters.
Three of these characterize the lens--source geometry: the time of closest approach $t_0$, 
the impact parameter at that time $u_0$, and the event timescale $t_{\rm E}$. Two additional
parameters describe the binary nature of the lens, which are the projected separation $s$ and 
the mass ratio $q \equiv M_2/M_1$ between the two lens components, where $M_1$ denotes the
component lying closer to the source trajectory. The parameter $\alpha$ specifies the angle
between the source trajectory and the binary axis. The parameters $u_0$ and $s$ are normalized
to the angular Einstein radius $\theta_{\rm E}$. The final parameter is the normalized source
radius $\rho \equiv \theta_*/\theta_{\rm E}$, where $\theta_*$ is the angular radius of the
source star. This parameter is required to model caustic-crossing features, which are strongly
affected by finite-source effects.

With the magnification computed from these parameters, the model flux is expressed as 
$F = A F_0 + F_b$, where $F_0$ is the unmagnified source flux and $F_b$ represents the
blended flux.  Blended light is explicitly accounted for in the modeling through a fitted 
blend flux parameter for each dataset.  Because the deviations are short-lived and occur at 
epochs and with morphologies predicted by caustic interactions in the lensing models, it is 
difficult to reproduce with intrinsic variability of a blended star.

Although the 2L1S models successfully reproduce the overall light curves for all events, subtle
residual features remain that cannot be explained within this framework. Such anomalies are
commonly interpreted as signatures of an additional component either in the lens system or in 
the source system. Motivated by this possibility, we therefore carried out additional modeling 
that incorporates an extra lens or source component.

When an additional source is introduced, the lens-system configuration corresponds to a 2L2S 
system. Describing the lensing behavior in this configuration requires several additional 
parameters beyond those of the 2L1S model. These include $(t_{0,2}, u_{0,2})$, which describe 
the time of closest approach and impact parameter of the source companion ($S_2$) relative to 
the lens, as well as the flux ratio $q_F$ between the companion and the primary source.  In 
cases for which the companion source also crosses or approaches a caustic, it is necessary to 
introduce the normalized source radius of the companion, $\rho_2$. For clarity, we denote 
the parameters associated with the primary source as $(t_{0,1}, u_{0,1})$ to distinguish 
them from those of the source companion.  We note that the possibility of an unrelated background 
source is highly unlikely given the low probability of such chance alignments and the absence of 
known stellar systems along the line of sight.

When an additional lens component is included, the lensing system is described by a triple-lens 
single-source (3L1S) configuration. As in the 2L2S case, modeling a 3L1S system requires additional 
parameters. These include $(s_3, q_3)$, which denote the projected separation and mass ratio of 
the third lens component ($M_3$) relative to the primary lens ($M_1$), and $\psi$, the orientation 
angle of $M_3$ measured with respect to the $M_1$–$M_2$ axis centered on $M_1$. A summary of the 
lens parameters adopted for different lens-system configurations is given in Table~2 
of \citet{Han2023}. In the following subsections, we present detailed analyses of the individual 
events.

In the analysis, we tested higher-order effects such as parallax, lens orbital motion, and xallarap. 
However, models incorporating these effects neither account for the residuals from the 2L1S model 
nor yield statistically significant improvements in the fit.

\subsection{KMT-2021-BLG-0209 \label{sec:three-one}}

The light curve of KMT-2021-BLG-0209 is shown in Figure~\ref{fig:one}. It displays the 
characteristic anomaly pattern of a caustic-crossing binary-lens event, featuring a weak 
bump centered around ${\rm HJD}^\prime \equiv {\rm HJD} - 2450000 \simeq 9305$, which is 
produced by the source's approach to a caustic cusp, and a pair of sharp spikes at 
${\rm HJD}^\prime \simeq 9318.13$ ($t_1$) and $\simeq 9319.20$ ($t_3$), corresponding to 
the source's entrance into and exit from the caustic.  The event was detected by the KMTNet 
team on March 31, 2021 (${\rm HJD}^\prime = 9304$), approximately one day before the first 
bump, and was observed exclusively by KMTNet. The source is located in the overlapping region 
of the two KMTNet prime fields BLG01 and BLG41, which were monitored with cadences of 0.5 hr 
for each field, resulting in an effective combined cadence of 0.25 hr.

\begin{figure*}[t]
\centering
\includegraphics[width=13.0cm]{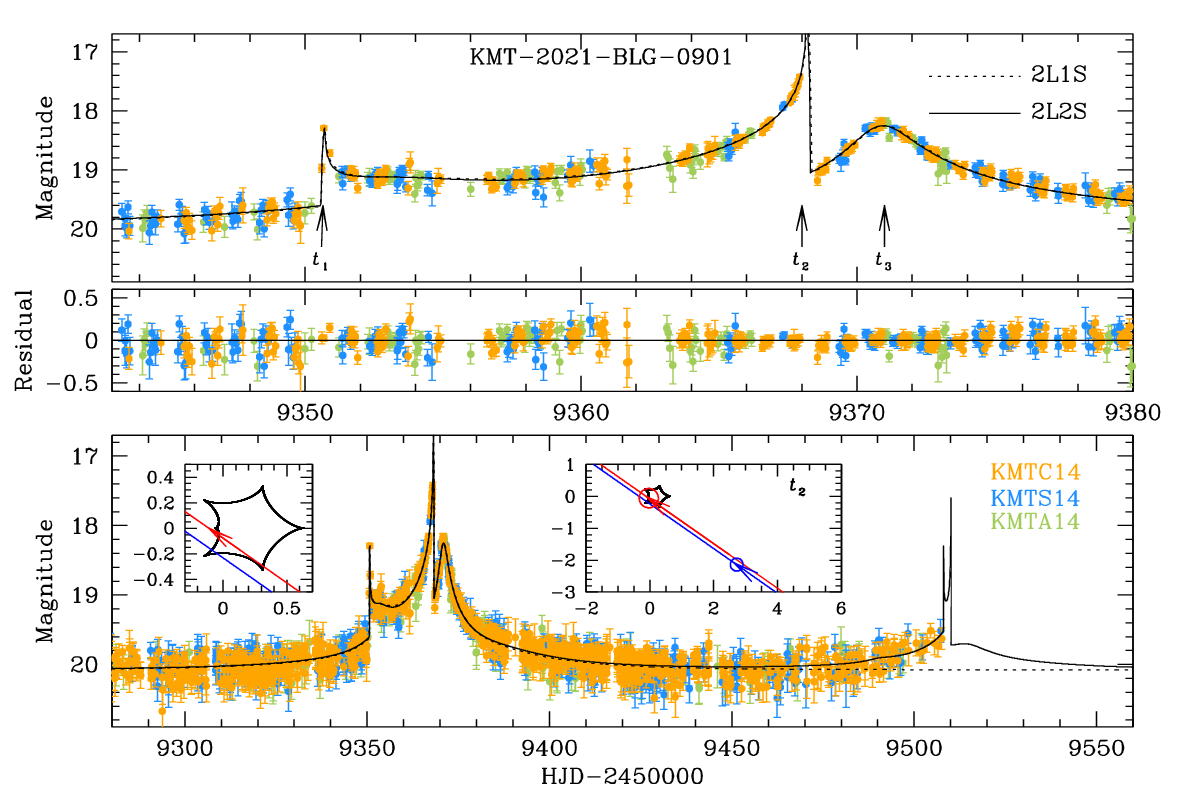}
\caption{
Lensing light curve, system configurations, and model curves of the lensing event 
KMT-2021-BLG-0901.  Notations are similar to those of Fig.~\ref{fig:one}.  The two 
insets in the lower panel show the lens-system configuration. The left inset presents 
a zoomed-in view of the caustic, while the right inset provides a zoomed-out view 
that includes the positions of the binary source stars $S_1$ and $S_2$ at time $t_2$.
}
\label{fig:two}
\end{figure*}

Modeling with a 2L1S configuration yields a solution that approximately reproduces the 
main features of the light curve. The event timescale is $t_{\rm E}\sim 85$~days, and the 
best-fit binary-lens parameters are $(s,q)\sim (3.2,1.3)$.  In Figure~\ref{fig:one}, the 
corresponding model is overplotted on the data as a dotted line.  According to this model, 
the binary lens produces two four-cusped caustics, each located near one of the lens 
components.  The weak bump at ${\rm HJD}^\prime \sim 9305$ arises from the source's close 
approach to the on-axis cusp of the caustic associated with the lower-mass lens component, 
while the caustic-crossing spikes are generated as the source traverses the lower portion 
of this caustic.

However, closer inspection reveals subtle deviations from the 2L1S model in the caustic-exit 
region, as shown in the upper panel of Figure~\ref{fig:one}.  In particular, a weak additional 
caustic-like feature appears, and the region between this feature and the major peak at $t_3$ 
shows systematic residuals relative to the 2L1S model. A qualitatively similar deviation was 
reported for the microlensing event OGLE-2016-BLG-1003, for which the anomaly was interpreted 
as arising from a caustic crossing of a binary source companion \citep{Jung2017}.  Motivated 
by this resemblance, we therefore explored an additional 2L2S model to examine whether it can 
better account for the observed features.  We also tested a 3L1S model to examine whether the 
residuals from the 2L1S fit could be explained by the presence of an additional lens component. 
We find that this model fails to reproduce the observed deviations.

The 2L2S modeling was performed by starting from the best-fit 2L1S solution and exploring 
a range of possible trajectories for the secondary source. Through this procedure, we 
identified a 2L2S solution that successfully accounts for all of the observed anomaly 
features, including those that cannot be explained by the 2L1S model. The complete set of 
2L2S model parameters is presented in Table~\ref{table:three}, and the corresponding best-fit 
light curve is overplotted on the data as a solid line in Figure~\ref{fig:one}.

The lens-system configuration of the best-fit 2L2S model is illustrated in the two insets 
in the lower panel of Figure~\ref{fig:one}. The left inset shows the overall caustic 
structure, while the right inset presents an enlarged view of the region where the source 
crosses the caustics.  The secondary source is relatively faint, with a flux ratio with 
respect to the primary source of $q_F \sim 0.075$.  The two source stars are closely spaced, 
with a normalized separation $\Delta u = \sqrt{(\Delta t_0/\te)^2 + \Delta u_0^2 )} = 0.0063$, 
and the secondary source trails the primary source by a time offset of $\Delta t_0 = 3.5$ hr.  
The secondary source enters the caustic at ${\rm HJD}^\prime = 9319.12$ ($t_2$), which occurs 
before the caustic exit of the primary source at $t_3$, and exits the caustic at 
${\rm HJD}^\prime = 9319.49$ ($t_4$), which is after $t_3$. These caustic crossings of $S_2$ 
produce residuals that could not be explained by a 2L1S model. The caustic entrance of the 
secondary source is expected to have produced a weak caustic feature, but this feature was 
not covered by the available data.

\subsection{KMT-2021-BLG-0901 \label{sec:three-two}}

Figure~\ref{fig:two} displays the light curve of the microlensing event KMT-2021-BLG-0901. 
The event was detected by the KMTNet survey on 2021 May 19 (${\rm HJD}^\prime = 9353$). 
The source lies in the KMTNet field BLG14, which was monitored with a cadence of 1~hr. 
The event persisted throughout the 2021 bulge season, and the magnification remained 
elevated beyond the end of the observing season.

The first part of the light curve, i.e., prior to ${\rm HJD}^\prime \sim 9450$, exhibits 
the characteristic pattern of a caustic-crossing binary-lens event. It shows a pair of 
sharp spikes at ${\rm HJD}^\prime \simeq 9351$ ($t_1$) and $\simeq 9368$ ($t_2$) produced 
by source's caustic crossings, followed by a broad bump centered at ${\rm HJD}^\prime 
\simeq 9371$ ($t_3$) that arises from the source's approach to a caustic cusp.  At later 
times, however, the light curve deviates from this behavior and begins to rise again.  
Because the light curve had already returned to baseline at the start of the 2022 season, 
this observed behavior implies that the underlying profile peaked between the 2021 and 
2022 seasons.

\begin{figure*}[t]
\centering
\includegraphics[width=13.0cm]{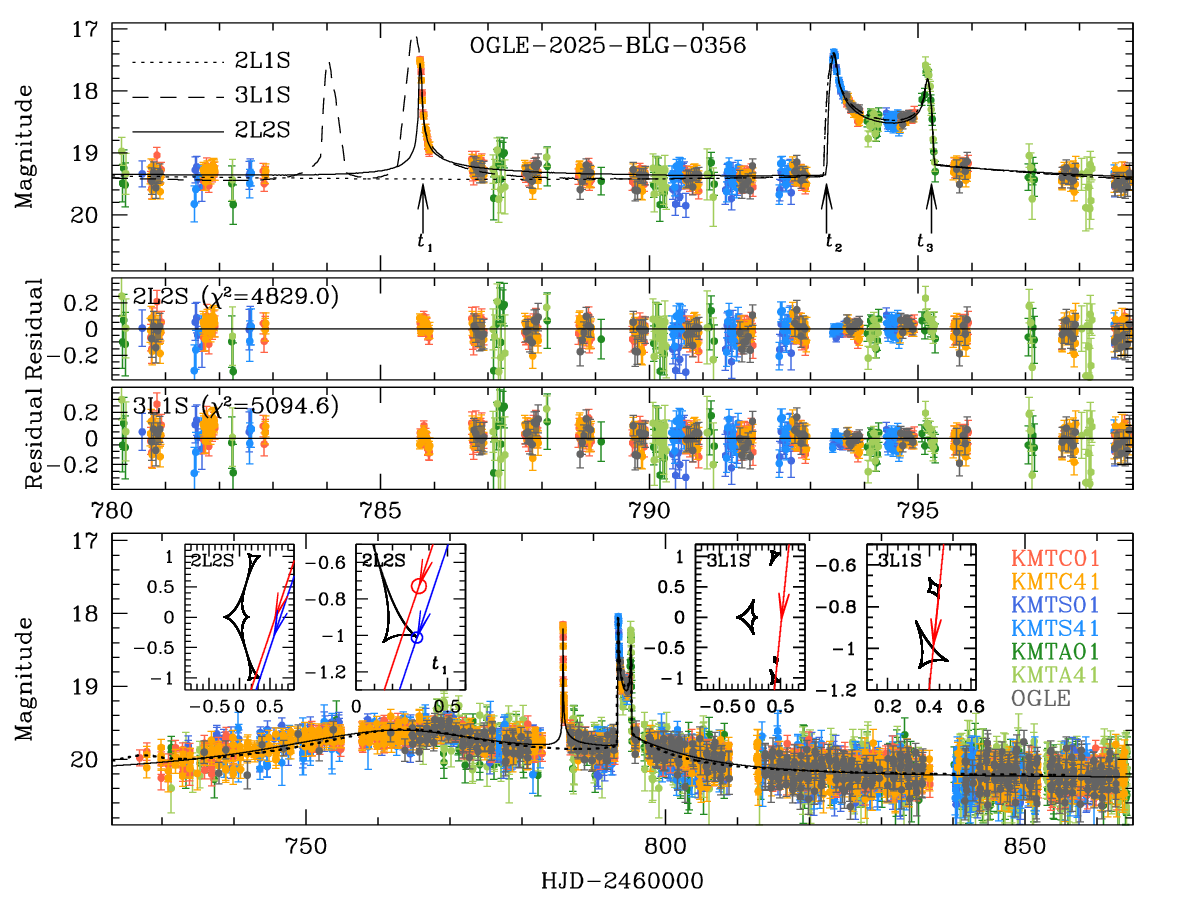}
\caption{
Light curve of the lensing event OGLE-2025-BLG-0356.  The lower panel shows two sets of 
lens-system configurations: the 2L2S configuration on the left and the 3L1S configuration 
on the right.  In the right 2L2S inset, the red and blue open circles indicate the positions 
of the binary source stars at $t_1$. The circles are not drawn to scale with respect to the 
source size.
}
\label{fig:three}
\end{figure*}

We first tested whether a 2L1S model could reproduce the full anomaly pattern, but found 
that this framework cannot adequately describe the light curve as a whole. The 2L1S model 
does, however, provide a satisfactory fit to the early portion of the event. For this 
solution, the best-fit binary-lens parameters are $(s, q) \sim (1.17, 0.12)$ and the 
event timescale is $t_{\rm E} \sim 39$~days. The corresponding 2L1S model curve is shown 
as a dotted line in Figure~\ref{fig:two}.

The light curve of KMT-2021-BLG-0901 resembles those of the 2L2S events KMT-2024-BLG-0412 
\citep{Han2024} and KMT-2024-BLG-2480 \citep{Han2025a} in that the main anomaly is well 
reproduced by a 2L1S model, whereas the late-time residuals are naturally explained by 
an additional source. Motivated by these similarities, we modeled the observed light curve 
using a 2L2S configuration. This modeling indicates that the event involves both a binary 
lens and a binary source. The full set of lensing parameters for the 2L2S solution is listed 
in Table~\ref{table:three}, and the corresponding model curve is overplotted on the data in 
Figure~\ref{fig:two} as a solid line.

The inset in the lower panel of Figure~\ref{fig:two} displays the lens-system configuration 
of the event.  The binary lens produces a resonant caustic composed of six folds.  The 
primary source crosses the caustic by entering through the lower-right fold and exiting 
through the lower-left fold, producing the sharp spikes at $t_1$ and $t_2$.  After the 
caustic exit, the source approaches the left on-axis cusp, generating the broad bump 
centered at $t_3$.  The secondary source is fainter than the primary by $\sim 0.92$ mag 
in the $I$ band and trails the primary with a time offset of $\Delta t_0 \sim 143$ days. 
This source also encounters the caustic, passing near the lower-left tip of the caustic 
and producing an additional caustic-crossing feature around ${\rm HJD}^\prime \simeq 9510$.  
This feature was not covered because the observations ended with the close of the 2021 
bulge season.  We emphasize that the preference for the 2L2S model over interpretations based 
on higher-order effects is driven by the observed late-time rebrightening, rather than by the 
predicted caustic-crossing feature at the end of the season.

\begin{figure*}[t]
\centering
\includegraphics[width=12.5cm]{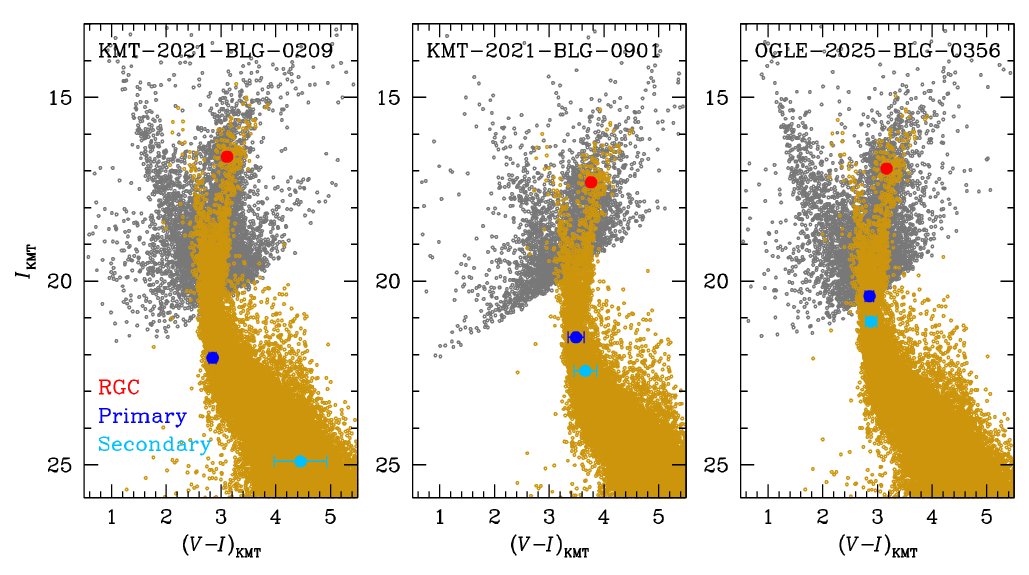}
\caption{
Locations of the binary-source stars in the color--magnitude diagram (CMD). The grey and brown 
CMDs are constructed from KMTC and HST observations, respectively.
}
\label{fig:four}
\end{figure*}

\subsection{OGLE-2025-BLG-0356 \label{sec:three-three}}

The lensing light curve of OGLE-2025-BLG-0356 is shown in Figure~\ref{fig:three}. It 
exhibits two prominent anomaly features. The first occurs near ${\rm HJD}^\prime \equiv 
{\rm HJD} - 2460000 = 785.8$ ($t_1$), and the second consists of a caustic-crossing 
structure with two sharp peaks at ${\rm HJD}^\prime \simeq 793.3$ ($t_2$) and ${\rm HJD}^\prime 
\simeq 795.3$ ($t_3$). The event was first identified and alerted by the OGLE group on 2025 
April 6 (${\rm HJD}^\prime = 771$), and was independently discovered by the KMTNet group, 
which issued a second alert on 2025 April 29 (${\rm HJD}^\prime = 794$). The source lies 
in the overlap region of the two KMTNet prime fields BLG01 and BLG41, which were monitored 
with a combined cadence of 15 minutes. The event was already in progress before the start 
of the 2025 bulge observing season.

Guided by the caustic-related anomaly features, we first modeled the event using a 2L1S 
configuration. This approach yields a solution that reproduces the overall light-curve 
morphology, including the sharp spikes at $t_2$ and $t_3$. For this model, the best-fit 
binary-lens parameters are $(s,q)\sim(0.66,2.7)$ with an event timescale of $t_{\rm E} 
\sim 30$~days. The corresponding 2L1S model is overplotted on the data as a dotted curve 
in Figure~\ref{fig:three}. The inferred binary lens produces three caustics: a central 
caustic located near the center of mass and two peripheral caustics displaced from it. 
In this interpretation, the features at $t_2$ and $t_3$ arise from the source crossing 
one of the peripheral caustics. However, the 2L1S model fails to account for the anomaly 
at $t_1$.

\begin{deluxetable}{lllllll}
\tablewidth{0pt}
\tablecaption{Source parameters and angular Einstein radius. \label{table:four}}
\tablehead{
\multicolumn{1}{c}{Parameter}           &
\multicolumn{1}{c}{KMT-2021-BLG-0209}   &
\multicolumn{1}{c}{KMT-2021-BLG-0901}   &
\multicolumn{1}{c}{OGLE-2025-BLG-0356}        
}
\startdata
 $(V-I)_{S_1}$                &  $2.847 \pm 0.062  $  &  $3.491 \pm 0.144  $  &  $2.854 \pm 0.079 $  \\
 $I_{S_1}$                    &  $22.091 \pm 0.014)$  &  $21.530 \pm 0.151 $  &  $20.409 \pm 0.017$  \\
 $(V-I, I)_{S_2}$             &  $4.454 \pm 0.479  $  &  $3.659 \pm 0.207  $  &  $2.884 \pm 0.098 $  \\
 $I_{S_2}$                    &  $24.907 \pm 0.188 $  &  $22.446 \pm 0.152 $  &  $21.101 \pm 0.027$  \\
 $(V-I, I)_{\rm RGC}$         &  $(3.111, 16.617)  $  &  $(3.764, 17.314)  $  &  $(3.166, 16.938) $  \\
 $(V-I, I)_{{\rm RGC},0}$     &  $(1.060, 14.501)  $  &  $(1.060, 14.446)  $  &  $(1.060, 14.475) $  \\
 $(V-I)_{S_1, 0}$             &  $0.796 \pm 0.062  $  &  $0.787 \pm 0.144  $  &  $0.748 \pm 0.079 $  \\
 $I_{S_1, 0}$                 &  $19.976 \pm 0.014 $  &  $18.661 \pm 0.151 $  &  $17.947 \pm 0.017$  \\
 $(V-I)_{S_2, 0}$             &  $(2.403 \pm 0.479,$  &  $(0.955 \pm 0.207 $  &  $0.778 \pm 0.098 $  \\
 $I_{S_2, 0}$                 &  $22.759 \pm 0.188)$  &  $19.578 \pm 0.152)$  &  $18.638 \pm 0.027$  \\
  Type ($S_1$)                &   G8V                 &   G8V                 &   G6V                \\
  Type ($S_2$)                &   M3V                 &   K2V                 &   G8V                \\
 $\theta_{*, S_1}$ ($\mu$as)  &  $0.350 \pm 0.033  $  &  $0.635 \pm 0.102  $  &  $0.846 \pm 0.090 $  \\
 $\theta_{*, S_2}$ ($\mu$as)  &  $0.272 \pm 0.132  $  &  $0.505 \pm 0.110  $  &  $0.636 \pm 0.076 $  \\
 $\thetae$ (mas)              &  $0.841 \pm 0.081  $  &  $0.381 \pm 0.079  $  &  $0.371 \pm 0.043 $  \\
 $\mu$ (mas/yr)               &  $3.62 \pm 0.35    $  &  $3.36 \pm 0.7     $  &  $4.07 \pm 0.47   $  
\enddata
\end{deluxetable}

A short-duration anomaly that cannot be explained by a 2L1S model may be produced by a 
third lens component, such as a planetary companion in a binary-lens system, as illustrated 
by the event KMT-2016-BLG-1337 \citep{Han2025b}. We therefore explored a 3L1S interpretation 
to examine whether the anomaly near $t_1$ can be explained by introducing a third lens 
component. In this modeling, we performed a grid search over the third-lens parameters 
$(s_3, q_3, \psi)$, while optimizing the remaining parameters through $\chi^2$ minimization, 
with initial values adopted from the 2L1S solution.

The lensing parameters of the best-fit 3L1S model are listed in Table~\ref{table:three}, 
and the corresponding model light curve is overplotted on the data in Figure~\ref{fig:three}. 
This model reproduces the overall morphology of the event, including the anomaly near $t_1$. 
The parameters associated with the third lens component are $(s_3, q_3)\sim (1.8, 7.5\times 
10^{-3})$, indicating that the third body is a planetary-mass companion located outside 
the Einstein ring. In this interpretation, the third body generates a small caustic along 
the source trajectory, and the anomaly around $t_1$ is attributed to the perturbation 
produced during the source passage near this caustic, as illustrated in the two right 
insets of Figure~\ref{fig:three}.

Because a short-term anomaly can also arise from a companion to the source, as in 
KMT-2019-BLG-1715 \citep{Han2021}, we also modeled the event using a 2L2S configuration. 
The best-fit 2L2S parameters are listed in Table~\ref{table:three}, and the model is shown 
in Figure~\ref{fig:three} as a solid line. This solution explains all anomaly features and 
improves the fit relative to the 3L1S model by $\Delta\chi^2 = 107.5$. We therefore identify 
OGLE-2025-BLG-0356 as a 2L2S event.

The lens-system configuration for the 2L2S solution is shown in the left insets of 
Figure~\ref{fig:three}. The primary source passes one of the two peripheral caustics 
induced by the binary lens, producing the caustic-crossing spike features at $t_2$ and 
$t_3$. The secondary source is fainter than the primary source by $\Delta I = -2.5 \log q_F 
= 0.69$ mag and leads the primary by $\Delta t_0 = 9.04$ days. It passes slightly outside 
the caustic, producing the anomaly feature at $t_1$.

\section{Source stars and Einstein radius \label{sec:four}}

In this section, we characterize the stars that constitute the binary source of the event. 
This characterization is also essential for determining the angular Einstein radius, which 
is given by
\begin{equation}
\theta_{\rm E} = \frac{\theta_*}{\rho}.
\label{eq2}
\end{equation}
\hskip-4pt
Here $\theta_*$ is the angular radius of the source star and $\rho$ is the normalized 
source radius obtained from the light-curve modeling. Therefore, estimating $\thetae$ 
requires a determination of $\theta_*$, which can be inferred from the source type.

To estimate the angular source radius, we follow the method of \citet{Yoo2004}. In this 
approach, the instrumental source color and magnitude, $(V-I, I)$, are first determined 
by regressing the observed light curve against the best-fit model. The de-reddened source 
color and magnitude, $(V-I, I)_0$, are then obtained by calibrating the source relative 
to a reference. For this calibration, we adopt the centroid of the red giant clump (RGC) 
in the instrumental color--magnitude diagram as the reference and use its intrinsic color 
and magnitude as determined by \citet{Bensby2013} and \citet{Nataf2013}. Finally, the 
angular source radius is subsequently inferred from empirical color--surface-brightness 
relation of \citet{Kervella2004}.

For some events, the source color could not be reliably measured because of poor 
photometric quality and sparse coverage of the $V$-band light curve. In such cases, 
we estimated the source color using a color--magnitude diagram constructed from Hubble 
Space Telescope observations \citep{Holtzman1998}. Specifically, we selected stars 
whose $I$-band magnitudes have the same offset from the red giant clump (RGC) centroid 
as the source and adopted the mean color of these stars as a proxy for the source color. 
The color of the secondary source was estimated in the same manner, using the flux ratio 
$q_F$ derived from the light-curve modeling.

In Figure~\ref{fig:four}, we present the positions of the individual binary-source stars 
on color--magnitude diagrams constructed by combining KMTC and Hubble Space Telescope 
observations.  Table~\ref{table:four} lists the instrumental colors and magnitudes of the 
primary and secondary sources, $(V-I, I)_{S_1}$ and $(V-I, I)_{S_2}$, together with those 
of the RGC centroid, $(V-I, I)_{\rm RGC}$. The table also provides the de-reddened colors 
and magnitudes of the primary and secondary sources, $(V-I, I)_{S_1,0}$ and $(V-I, I)_{S_2,0}$, 
as well as those of the RGC centroid, $(V-I, I)_{{\rm RGC},0}$, and the inferred spectral 
types of the source stars. In addition, we list the angular Einstein radius computed from 
Equation~(\ref{eq2}) and the relative lens--source proper motion, $\mu = \thetae/\te$. We 
find that the binary sources are composed of two main-sequence stars, with spectral types 
(G8V, M3V) for KMT-2021-BLG-0209, (G8V, K2V) for KMT-2021-BLG-0901, and (G6V, G8V) for 
OGLE-2025-BLG-0356.

Direct spectroscopic confirmation of the inferred spectral types would be challenging 
with current facilities due to the faintness of the sources. However, future extremely 
large telescopes may enable high-resolution spectroscopy or imaging that can disentangle 
the source components and provide independent constraints on their spectral types.

\begin{deluxetable}{lllllll}
\tablewidth{0pt}
\tablecaption{Physical lens parameters. \label{table:five}}
\tablehead{
\multicolumn{1}{c}{Parameter}           &
\multicolumn{1}{c}{KMT-2021-BLG-0209}   &
\multicolumn{1}{c}{KMT-2021-BLG-0901}   &
\multicolumn{1}{c}{OGLE-2025-BLG-0356}        
}
\startdata
 $M_1$ ($M_\odot$)     &  $0.72^{+0.37}_{-0.34}$  &  $0.56^{+0.30}_{-0.30}   $  &  $0.28^{+0.30}_{-0.15}$  \\ [0.5ex]
 $M_2$ ($M_\odot$)     &  $0.53^{+0.27}_{-0.25}$  &  $0.059^{+0.032}_{-0.032}$  &  $0.56^{+0.59}_{-0.30}$  \\ [0.5ex]
 $\dl$ (kpc)           &  $5.51^{+1.13}_{-1.52}$  &  $7.02^{+0.89}_{-1.41}   $  &  $7.52^{+1.05}_{-1.19}$  \\ [0.5ex]
 $a_\perp$ (au)        &  $16.6^{+2.7}_{-4.3}  $  &  $3.28^{+0.42}_{-0.66}   $  &  $2.19^{+0.31}_{-0.35}$  \\ [0.5ex]
 $p_{\rm disk}$        &  $70\%                $  &  $34\%                   $  &  $24\%                $  \\ [0.5ex]
 $p_{\rm bulge}$       &  $30\%                $  &  $66\%                   $  &  $76\%                $  \\ [0.5ex]
\enddata
\end{deluxetable}

\begin{figure}[t]
\includegraphics[width=\columnwidth]{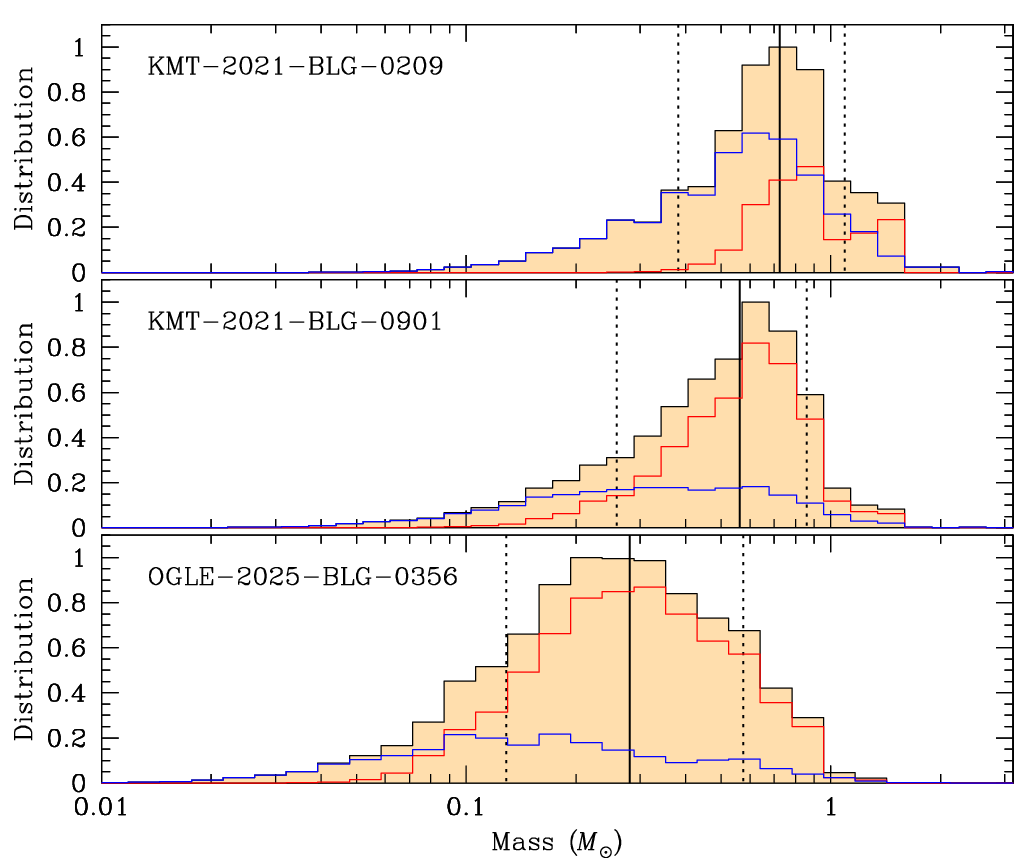}
\caption{
Bayesian posterior distribution of the primary lens mass.  The blue and red curves 
show the contributions from the disk and bulge lens populations, respectively. The 
solid vertical line indicates the median of the posterior, and the dotted lines 
indicate the $1\sigma$ range.
}
\label{fig:five}
\end{figure}

\section{Lens parameters \label{sec:five}}

We constrain the physical properties of the binary lens by estimating its mass and distance. 
To this end, we perform a Bayesian inference analysis based on the measured microlensing 
observables $\te$ and $\thetae$, which are related to the lens mass ($M$) and distance 
($\dl$) through
\begin{equation}
\te = {\thetae \over \mu};\qquad
\thetae = \sqrt{\kappa M \pi_{\rm rel}}.
\label{eq3}
\end{equation}
\hskip-4pt
Here $\kappa = 4G/(c^2{\rm au}) \simeq 8.14~{\rm mas}/M_\odot$, $\pi_{\rm rel}={\rm au}
(1/\ds-1/\dl)$ denotes the relative parallax of between the lens and source, and $\ds$ 
is the distance to the source.  The Bayesian analysis combines the measured values of 
the observables with prior knowledge of the Galactic populations that can act as lenses 
and sources, including their spatial distributions and kinematics. Within this framework, 
the posterior probability distribution for the physical parameters is obtained as
\begin{equation}
P(M, \dl ~|~ \te, \thetae) \propto P(\te, \thetae ~|~ M, \dl) P(M, \dl),
\label{eq4}
\end{equation}
\hskip-4pt
where $P(\te, \thetae ~|~ M, \dl)$ is the likelihood describing the consistency between 
a model defined by $(M, \dl)$ and the observed $\te$ and $\thetae$, and $P(M, \dl)$ 
denotes the prior probability density determined by the expected Galactic distributions 
of lenses and sources.

\begin{figure}[t]
\includegraphics[width=\columnwidth]{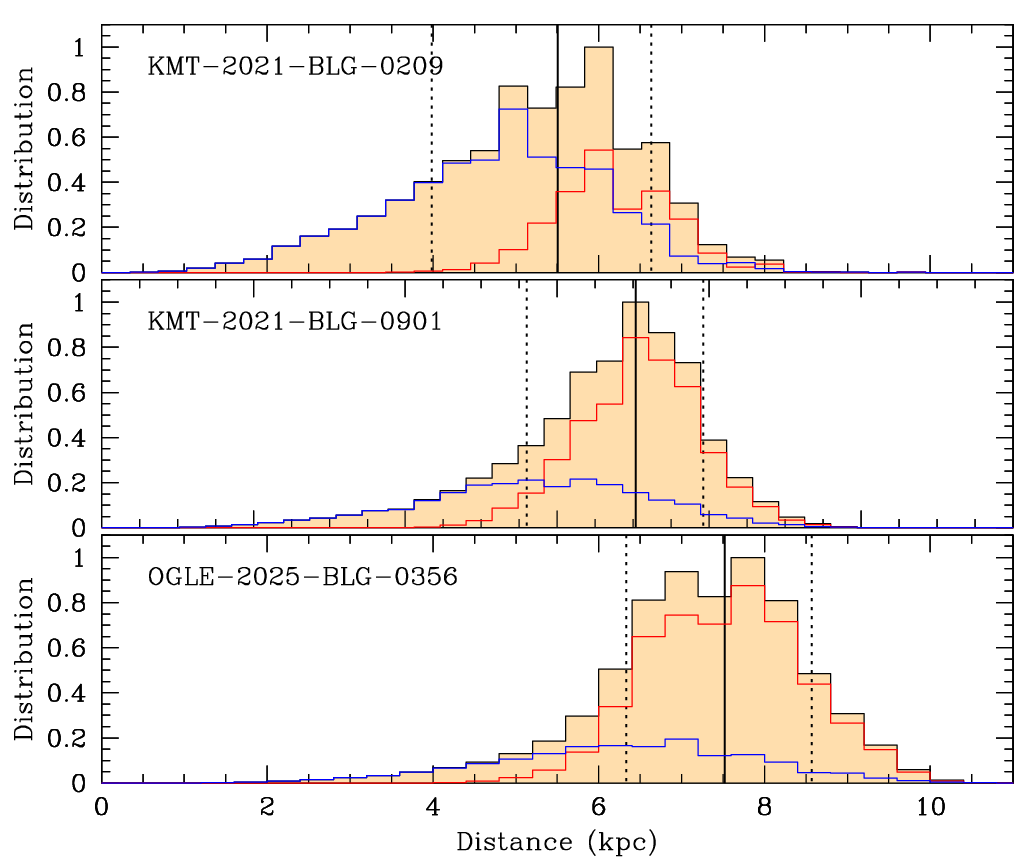}
\caption{
Bayesian posterior distribution of the distance to the lens mass.
}
\label{fig:six}
\end{figure}

The Bayesian priors were constructed using a Galactic framework that incorporates the 
density and kinematic distributions of disk and bulge stars, together with an assumed 
mass function for the lens population.  In the analysis, we adopted the Galactic model 
described by \citet{Jung2021} and used the lens mass function introduced by 
\citet{Jung2022}.  Based on these priors, we generated artificial lensing events from 
a large ensemble of lens--source pairs via Monte Carlo simulations. For each simulated 
event, the microlensing observables $\te$ and $\thetae$ were computed using 
Equation~(\ref{eq3}), and a relative likelihood was assigned according to 
\begin{equation}
L \propto \exp \left( -{\chi^2 \over 2}\right),
\label{eq5}
\end{equation}
\hskip-4pt
where
\begin{equation}
\chi^2 =
\frac{(t_{\rm E}-t_{\rm E,obs})^{2}}{\sigma_{t_{\rm E}}^2} +
\frac{(\theta_{\rm E}-\theta_{\rm E,obs})^{2}}{\sigma_{\theta_{\rm E}}^2}.
\label{eq6}
\end{equation}
\hskip-4pt
Here, $(t_{\rm E,obs}, \theta_{\rm E,obs})$ are the measured values of the microlensing 
observables, and $(\sigma_{t_{\rm E}}, \sigma_{\theta_{\rm E}})$ denote the corresponding 
uncertainties.  We note that $\thetae$ is derived from the normalized source radius $\rho$ 
and the amgular source radius $\theta_*$ is independently estimated from the source color 
and magnitude.  Possible correlations between $\te$ and $\thetae$ arising from the light-curve 
modeling are not explicitly included in the likelihood, but are expected to have a negligible 
impact on the inferred physical parameters.

The resulting posterior probability distributions are presented in Figure~\ref{fig:five} 
for the lens mass and in Figure~\ref{fig:six} for the lens distance.  The inferred physical 
properties of the lens system are summarized in Table~\ref{table:five}, where $a_\perp$ 
denotes the projected separation between the lens components.  As a representative estimate, 
we adopted the median of each posterior distribution, with uncertainties defined by the 16th 
and 84th percentiles. We also report the probabilities that the lens is located in the Galactic 
disk ($p_{\rm disk}$) or in the bulge ($p_{\rm bulge}$).

Based on their estimated masses, the lenses of KMT-2021-BLG-0209 and OGLE-2025-BLG-0356 
are binary systems composed of sub-solar-mass stars, which dominate the lens population 
in Galactic microlensing events \citep{Han2003}. In contrast, although the primary lens 
of KMT-2021-BLG-0901 is also a low-mass star, its companion falls in the brown-dwarf 
regime owing to the low mass ratio ($q \sim 0.1$). The lenses of KMT-2021-BLG-0901 and 
OGLE-2025-BLG-0356 are more likely located in the Galactic bulge, whereas KMT-2021-BLG-0209 
has a higher probability of residing in the disk.

\section{Summary and conclusion \label{sec:six}}

In this paper, we analyzed three anomalous microlensing events, KMT-2021-BLG-0209, 
KMT-2021-BLG-0901, and OGLE-2025-BLG-0356, whose light curves cannot be fully explained 
by the standard binary-lens single-source model. These events were identified through a 
systematic re-examination of anomalous KMTNet survey light curves for which earlier 
modeling attempts either failed or left subtle but persistent residuals.

We showed that the observed deviations can be described by four-body configurations 
that required an additional source component. For KMT-2021-BLG-0209, weak caustic-exit 
deviations were naturally reproduced by a faint companion source undergoing an additional 
caustic interaction. For KMT-2021-BLG-0901, the late-time re-brightening was explained by 
a second magnification episode that occurred when the secondary source encountered the 
caustic long after the primary. For OGLE-2025-BLG-0356, we tested the degeneracy between 
the 3L1S and 2L2S interpretations of the short, isolated anomaly and found that the 2L2S 
solution provided a substantially better fit. This result demonstrated that brief anomalies 
were not uniquely indicative of a third lens component.

These results highlight that as survey cadence and photometric precision improve, 
subtle deviations requiring higher-order modeling will become increasingly common. 
Routine consideration of four-body models will therefore be essential both for robust 
interpretation of anomalies. This is especially important for the forthcoming Roman 
Space Telescope microlensing survey, where enhanced precision will increase sensitivity 
to weak anomalies and demand systematic testing of additional source and lens components 
for reliable physical inference.

\begin{acknowledgements}
C.H. was supported by the Chungbuk National University 2025 NUDP program and the National 
Research Foundation of Korea (RS-2025-21073000).
This research has made use of the KMTNet system operated by the Korea Astronomy and Space 
Science Institute (KASI) at three host sites of CTIO in Chile, SAAO in South Africa, and 
SSO in Australia. Data transfer from the host site to KASI was supported by the Korea 
Research Environment Open NETwork (KREONET). This research was supported by KASI under 
the R\&D program (project No. 2026-1-904-01) supervised by the Ministry of Science and ICT.
W.Z. and H.Y. acknowledge support by the National Natural Science Foundation of China 
  (Grant No. 12133005). 
H.Y. acknowledges support by the China Postdoctoral Science Foundation (No. 2024M762938). 
W.Zang acknowledges the support from the Harvard-Smithsonian Center for Astrophysics 
  through the CfA Fellowship. 
J.C.Y. acknowledges support from U.S. NASA Grant No. 80NSSC25K7146. 
J.C.Y. acknowledges support from a Scholarly Studies grant from the Smithsonian Institution.
The OGLE project has received funding from the Polish National Science
Centre grant OPUS-28 2024/55/B/ST9/00447 awarded to AU.
H.Y. and W.Z. acknowledge support by the National Natural Science Foundation of China (Grant No. 12133005). H.Y. acknowledge support by the China Postdoctoral Science Foundation (No. 2024M762938).  
\end{acknowledgements}


\bibliography{references}
\bibliographystyle{aasjournalv7}

\end{document}